\documentclass[conference]{IEEEtran}
\IEEEoverridecommandlockouts
\usepackage{amsmath,amssymb,amsfonts}
\usepackage{xcolor,float}
\usepackage{booktabs,cite}
\usepackage{amsmath,amssymb,amsfonts}
\usepackage{graphicx,textcomp,url}
\begin{document}
\title{Innovative Platform for~Designing\\Hybrid Collaborative \& Context-Aware \\Data Mining Scenarios} 
\author{
\IEEEauthorblockN{Anca Avram, Oliviu Matei, Camelia-M. Pintea, Carmen Anton}
\IEEEauthorblockA{\textit{Technical University Cluj-Napoca,Romania} \\
anca.avram@ieee.org, oliviu.matei@holisun.com,\\dr.camelia.pintea@ieee.org, carmen.anton@cunbm.utcluj.ro}
}
\maketitle
\begin{abstract}The process of knowledge discovery involves nowadays a major number of techniques. Context-Aware Data Mining (CADM) and Collaborative Data Mining (CDM) are some of the recent ones. The current research proposes a new hybrid and efficient tool to design prediction models called Scenarios Platform-Collaborative \& Context-Aware Data Mining (SP-CCADM). Both CADM and~CDM approaches are included in~the~new platform in a flexible manner; SP-CCADM allows the setting and~testing of multiple configurable scenarios related to data mining at once. The introduced platform was successfully tested and validated on real life scenarios, providing better results than each standalone technique--CADM and CDM. Nevertheless, SP-CCADM was validated with various machine learning algorithms--k-Nearest Neighbour (k-NN), Deep Learning (DL), Gradient Boosted Trees (GBT) and Decision Trees (DT). SP-CCADM makes a step forward when confronting complex data, properly approaching data contexts and collaboration between data. Numerical experiments and~statistics illustrate in detail the potential of the proposed platform.
\end{abstract}
\begin{IEEEkeywords}{context-aware data mining; collaborative data mining; machine learning}
\end{IEEEkeywords}
\section{Introduction}
Nowadays, technology allows the storing of larger amounts of data. Having this data analyzed in a proper manner could help us enhance our processes and discover important patterns in data, that would lead to improvements in every domain this knowledge is applied to. 

Collecting data is a process that is still dependent on different sensors, programs or machines. Any disruption in the functioning of the data provider can result in~loss of data or noise in the obtained data. That is a reason why various approaches are the subject of continuous research in the data mining~processes.

Han et al.~\cite{han2011data} emphasize the~need to have different techniques for covering the~discrepancies that are brought in~the~data mining process by the incomplete, noisy or inconsistent data~\cite{cricsan2012risk}. 

Stahl~et~al.~\cite{stahl2010pocket} use the Pocket Data Mining term to define the collaborative mining of streaming data in mobile and distributed computing environments and propose an architecture in this direction.

Correia~et~al.~\cite{correia2010architecture} also designed a~collaborative framework allowing researchers to share the results and~their expertise so that these can be further used in other research. Web services were implemented and deployed and were responsible for seeking relevant knowledge among the collaborative web sites. They designed and~deployed a prototype for collaborative data mining in the fields of Molecular Biology and Chemoinformatics. In Reference~\cite{Fenza2011AHC}, data mining extract rules associate user profile and context features with an eligible set of recommendable points of interest to tourists. 

Matei et al.~\cite{matei2017multi,matei2017indin} proposed for the first time a multi-layered architecture for data mining in the context of Internet of Things (IoT), where a special place is~defined for context-aware, respective collaborative data mining. The concept takes into account the characteristics of the data, throughout its flow from the sensors to the cloud, where complex processing can be performed. At the local level, simple calculations can be performed usually due to the limitations imposed by the embedded systems or by the~communication infrastructure. In the cloud, the data mining goes from stand-alone algorithms, applied for one data source solely, to context-extraction and context-aware~\cite{weiser1999origins,bouquet2003c,voida2002integrating} approach and, finally, to collaborative processing, meaning the combination of more (correlated) data sources for improving the accuracy of analysis of one of them.

Previous research has proven that using collaborative data mining (CDM) and context-aware data mining (CADM) versus the classical data mining approach would lead to better results~\cite{avram2019context}.

The current study makes a step further and~extends the work performed in Reference \cite{anton2019performance} and analyzes how these two approaches would work in different scenarios for this matter, a new hybrid technique was considered, Scenarios Platform-Collaborative \& Context-Aware Data Mining (SP-CCADM), which would allow the~testing of more combinations and interactions between CADM and~CDM. The proposed model  was then applied and validated in a real-life scenario.

The remainder of the article is structured as follows---Section~\ref{subsection:methodCDM} introduces the~fundamentals of collaborative data mining. Section~\ref{subsection:methodCADM} presents the~concepts related to~context-aware data mining and~Section~\ref{subsection:cadm-cdm} introduces the~SP-CCADM technique. Section~\ref{section:methods} shows the~experimental setup, namely the~analysis technique, the~data sources, the~methods used and~the~implementation. Section~\ref{sec:results} illustrates both experimental results and~statistical analysis followed by disscusions, conclusions and~further work presented in~the~last part of the~research~paper.

\subsection{Collaborative Data Mining (CDM)} \label{subsection:methodCDM}

Collaborative data mining is~a~technique of approaching a~machine learning process that involves completing the~data of a~studied source with data taken from other similar sources~\cite{anton2019performance}. The~objective of the~process is~to provide better results than the~one that only uses the~data of the~studied~source. 

Mladenic~et~al.~\cite{mladenic2003data} and~Blokeel~et~al.~\cite{blockeel2002collaborative} performed experiments that used a collaborative data mining process between teams that share knowledge and~results.

A data collaboration system was implemented and~studied by Anton~et~al. in~Reference \cite{anton2019collaborative}. The~obtained results were compared with the~ones obtained using only the~data from a~single source. The~conclusion was that adapting the~used algorithms and~the~parameter setup for~these algorithms, can lead to~improved outcomes. Also, previous research performed by Matei~et~al. in~\cite{matei2016collaborative} has shown that the~accuracy of the~prediction increases with the~increase of the~data sources~correlation.

\subsection{Context Aware Data Mining (CADM)} \label{subsection:methodCADM}

Context-awareness became a~research subject starting from the~early 2000s (\cite{weiser1999origins,bouquet2003c,voida2002integrating}). 
According~to~the~definition by Dey~\cite{dey2001understanding}, context ``is any information that can be~used to~characterize the~situation of an entity. An~entity is~a~person, place, or~object that is~considered relevant to~the~interaction between a~user and~an application, including the~user and~applications themselves.'' Lee~et~al.~\cite{lee2011survey} say that a~context-aware system is~one that could adapt its operations actively using the~existing contextual~information. 

Context aware data mining, beside the~classical data mining approach comes with an extra step of integrating context data in~the~process. Lee~et~al.~\cite{lee2011survey}, identified the~phases of context-aware data mining as being---(1) Acquisition of context (usually performed with the~use of different physical or virtual sensors~\cite{yang2006context}); (2) Storage of context (in files, databases, repositories depending on the~data characteristics); (3) Knowledge analysis, where context is~either aggregated, or~elevated on the~level of semantics describing the~data; (4) Use of context~data.

{The research performed by Stokic~et~al.~\cite{stokic2014generic} specifies that context sensitivity can enhance the~observation of the~operating parameters for~a~system. The~conclusion is~that systems could dynamically adjust when scenarios~change.}

Scholze~et~al.~\cite{scholze2017holistic} identified context sensitivity as a~reliable option to~create a~holistic solution for~(self-)optimization of discrete flexible manufacturing systems. 
Perera~et~al.~\cite{perera2013context} conducted an extensive survey on the~context aware computing efforts in~the~IoT. They concluded that context awareness is~of main importance and~understanding sensor data is~one of the~biggest challenges in~the~IoT.

Scholze~et~al.~\cite{scholze2013context} made the~proposal of using context awareness to~implement context-sensitive decision support services in~an eco-process engineering system setting. Vajirkar~et~al.~\cite{vajirkar2003context} identified the~advantages of using CADM for~wireless devices in~the~medical field and~proposed a~CADM framework to~test the~suitability of different context~factors.

\subsection{Combining CADM and~CDM in~a~Flexible~Architecture}\label{subsection:cadm-cdm}
The quality of the~information available for~analysis is~very important in~the~knowledge discovery process. As~Marakas emphasizes~\cite{marakas2003modern}, this ``can make or break the~data mining~effort''.

\begin{figure*}[htbp]
\centering
\includegraphics[scale=0.5]{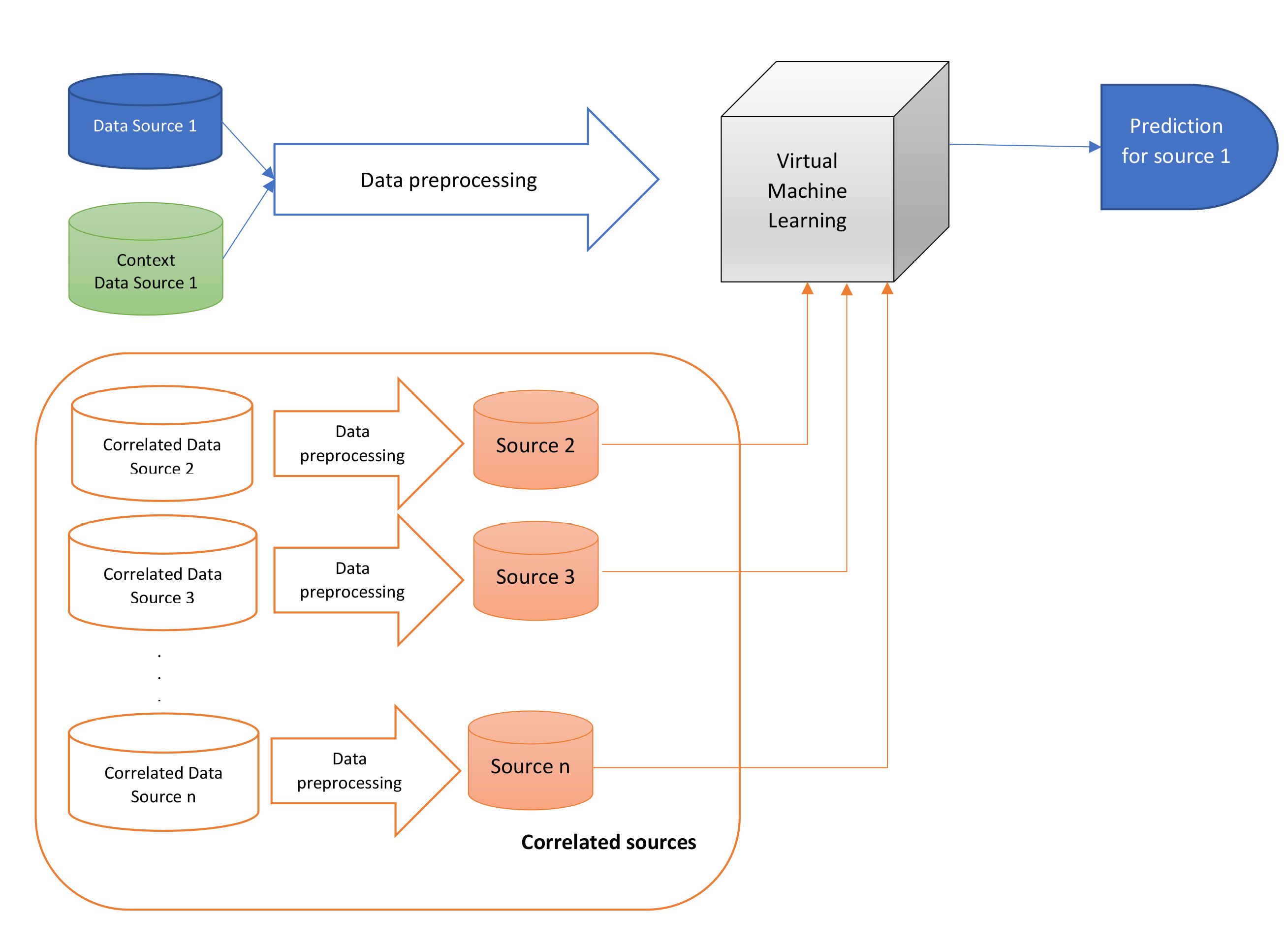}
\caption{Innovation: Context-Aware Data Mining (CADM) and~Collaborative Data Mining (CDM) combined process overview, influenced by Reference~\cite{anton2019performance}.}
 	\label{fig:combinedProcessesDiagram}
 \end{figure*}

The previous work~\cite{anton2019performance} concluded that both CADM and~CDM techniques offer advantages against the~classical data mining approach; the~current work  makes a~step forward and~provides a~hybrid approach of CADM and~CDM as depicted in~Figure~\ref{fig:combinedProcessesDiagram}.

The decision on what information to~use as context and~what data can be~used in~a~collaborative data mining environment depends very much on the~experience of the~person performing the~analysis. Information that could be~of use in~a~scenario, could have less value in~another situation. Also,~the~results may vary based on the~machine learning algorithms applied in~the~process. According~to~Ziafat and~Shakeri~\cite{ziafat2014using}, ``data mining algorithms are~powerful but cannot effectively work without the~active support of business~experts''.
The main purpose of this article is~to offer a~model of a~hybrid technique Scenarios Platform-Collaborative \& Context-Aware Data Mining (SP-CCADM) that would allow researchers to~easily test various combinations of CADM and~CDM with one or more collaborative sources, allowing them to~choose the~best possible scenario, based on the~obtained~results.
\section{Data and~Methods} \label{section:methods}
Section~\ref{sec:analysis_technique} presents an overview of the~SP-CCADM proposed technique: the~preconditions for~implementing, followed by a~detailed description. In~Section~\ref{sec:data_sources} data sources used for~the~proof of concept are~included, followed by the~methods (Section \ref{sec:methods}) and~implementation (Section \ref{sec:implementation}).
\subsection{Proposal: Scenarios
Platform-Collaborative \& Context-Aware Data Mining (SP-CCADM)}\label{sec:analysis_technique}

\subsubsection{Preliminary Analysis~Steps}\label{sec:prec}

\begin{itemize}
    \item Identify main data (MD) that is~the~subject of analysis, with~attributes $A_{M_1}$, $A_{M_2}$,...$A_{M_n}$. We~denote the~attribute that is~the~subject for~the~prediction with $A_{MP}$.
    \item Identify whether there is~a~possible suitable context that could be~used in~the~analyzed scenario. The~suite of k attributes corresponding to~the~context will be~noted with $A_{C_1}$, $A_{C_2}$,...$A_{C_k}$. 
    \item Identify possible collaborative sources ($CS_1$,$CS_2$, ... $CS_P$), each with a~variable $s_i$ number of attributes $A_{CS_j}$  that could be~used.
    \item Choose the~machine learning algorithms that seem suitable for~the~problem at hand.
    \item Decide upon the~measures that you would want to~measure when deciding on the~best possible~combinations.
    \item {Define the~test scenarios that you would want to~analyze.
    Table~\ref{table:testScenarios} defines an example of scenarios that could be~analysed. Question mark for~attribute name means that the~attribute is~not~ considered.}
\end{itemize}

\begin{table*}[htbp] 
\caption{Scenarios Platform-Collaborative \& Context-Aware Data Mining (SP-CCADM) example: a~hybrid CADM-CDM test scenarios to~be covered in~the~analysis, where $?$ are~ignored~attributes.}
\label{table:testScenarios}
\centering
\begin{tabular}{ccccc ccc ccccccc}
\toprule
\multicolumn{5}{c}{\textbf{Main Data}} & \multicolumn{3}{c}{\textbf{Context attributes}} & \multicolumn{7}{c}{\bf Collaborative Sources}\\\midrule
\multicolumn{5}{c}{\textbf{}} & \multicolumn{3}{c}{\textbf{}} & \multicolumn{3}{c}{\boldmath{$CS_1$}} & \textbf{….} & \multicolumn{3}{c}{\boldmath{$CS_P$}} \\ 
\midrule
\boldmath{$A_{M_1}$} & \textbf{…} & \boldmath{$A_{MP}$} & \textbf{…} & \boldmath{$A_{M_n}$} & \boldmath{$A_{C_1}$} & \textbf{…} & \boldmath{$A_{C_k}$} & \boldmath{$A_{CS_1}$} & \textbf{…} & \boldmath{$A_{CS_1}$} & \textbf{…} & \boldmath{$A_{CS_1}$} &  \textbf{…} & \boldmath{$A_{CS_j}$} \\ \midrule
val &   & val &  & val  & val &  & val & val &  & val  &  & val  &  & val  \\ 
val &   & val &  & val  & val &  & val & val &  & val  &  & ?  &  & ?  \\ 
val &   & val &  & val  & val &  & val & ? &  & ?  &  & ?  &  & ?  \\ 
val &   & val &  & val  & ? &  & ? & val &  & val  &  & val  &  & val  \\ 
val &   & val &  & val  & ? &  & ? & ? &  & ?  &  & val  &  & val  \\ 
val &   & val &  & val  & ? &  & ? & ? &  & ?  &  & ?  &  & ?  \\ \bottomrule
\end{tabular}
\end{table*}

\subsubsection{SP-CCADM Description on Data Mining~Algorithm} \label{sec:descriptionperAlg}

The hybrid data mining process has the~following~stages:

\begin{itemize}
    \item Load main data. 
    \item Load context data.
    \item Load correlated sources data.
    \item for~each defined test~scenario:
        \begin{itemize}
            \item Preprocess context data attributes specified in~the~test scenario; add it~to the~main data~source.
            \item Preprocess collaborative sources specified in~the~test scenarios and~add specified attributes to~the~main data source.
            \item Mark the~item specified in~the~test scenario as wanted prediction.
            \item Apply machine learning algorithm.
            \item Register chosen measure results for~the~chosen scenario.
        \end{itemize}
    \item In~the~end, analyze the~best scenario suitable for~the~chosen machine learning algorithm and~combination of CADM and~CDM.
\end{itemize}

\begin{figure*}[htbp]
\centering
\includegraphics[scale=0.8]{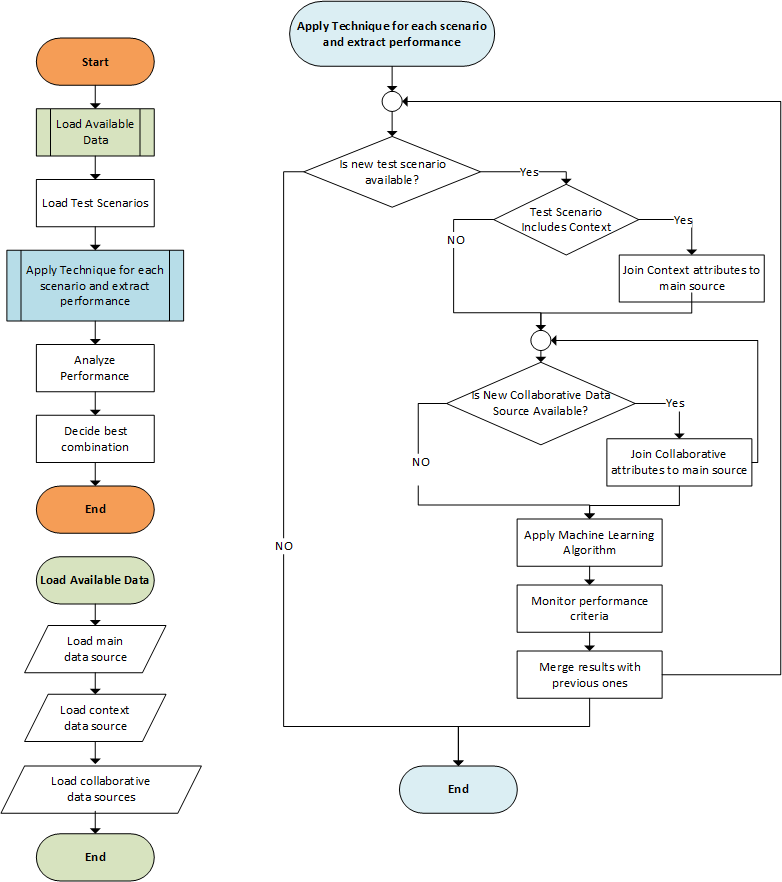}
\caption{Scenarios Platform-Collaborative \& Context-Aware Data Mining (SP-CCADM) flow~chart.}
 	\label{fig:flowChart}
 \end{figure*}

SP-CCADM is~illustrated in~the~flowchart diagram represented in~Figure~\ref{fig:flowChart}. Further on, the~article presents how the~ technique was used in~a~real life scenario for~predicting soil humidity for~a~location. 

\subsection{Data~Sources}\label{sec:data_sources}

Data used for~implementing the~proposed technique were downloaded from public sites that offer current weather prognosis, and~also allow access to~the~archived meteorological information gathered from weather stations around the~globe. 
Worldwide there are~different studies that rely on data offered by these sites. For~example, Vashenyuk~et~al.~\cite{vashenyuk2011study} used available data on precipitations to~study their relation to~radiations produced by thunderstorms. Siatnov~et~al.~\cite{sitnov2017link} used meteorogical data when trying to~explain the~link between the~2016 smoky atmosphere in~European Russia and~the~Siberian wildfires and~the~atmospheric~anomalies. 

Table~\ref{table:dataSources} presents an overview of collected data used in~the~experiments. The~first data set is~the~main one used in~the~experiments, while the~other is~a~control data set, used to~validate the~conclusions for~some specific scenarios. For~each location we~have one entry per observed day.

The~data series regarding the~soil moisture from the~six locations are~highly correlated, as~shown in~Table~\ref{table:correl} and~therefore seem to~be good candidates for~the~CDM~scenario.

\begin{table*}[htbp]
\caption{Overview of Data Sources with details about considered Time interval (time series), the~name of locations from Data Sources and~where to~find data on public~websites.}
\centering
\label{table:dataSources}
\begin{tabular}{llll}
\toprule
\textbf{Data Sources} & \textbf{Time Interval} &  \textbf{Locations} & \textbf{Public Data} \\ \bottomrule
\begin{tabular}[c]{@{}l@{}}6 locations in~Transylvania, \\ Romania\end{tabular} & 01.01.2016 to~31.12.2018 & \begin{tabular}[c]{@{}l@{}}Sarmasu, Reghin, Targu Mures, \\ Ludus, Blaj, Dumbraveni\end{tabular} & website~\cite{russianPrognosis}\\
\begin{tabular}[c]{@{}l@{}}4 locations in~Alberta Province, \\ Canada\end{tabular} & 01.05.2018 to~01.04.2020 & \begin{tabular}[c]{@{}l@{}}Breton, St. Albert, \\ Tomahawk, Leedale \end{tabular} & website~\cite{albertaWeather} \\ \bottomrule
\end{tabular}
\end{table*}
\unskip

\begin{table*}[htbp]
\caption{The correlation matrix of the~data sources from the~six locations~\cite{russianPrognosis}.}
\centering
\label{table:correl}
\begin{tabular}{lccccccc}
\toprule
 & \textbf{Campeni} & \textbf{Sarmasu} & \textbf{TMures} & \textbf{Reghin} & \textbf{Ludus} & \textbf{Blaj} & \textbf{Dumbraveni}\\ \midrule
Campeni & 1 & 0.751 & 0.743 & 0.651 & 0.729 & 0.785 & 0.741\\
Sarmasu & 0.751 & 1 & 0.902 & 0.880 & 0.931 & 0.867 & 0.858\\
TMures & 0.743 & 0.902 & 1 & 0.861 & 0.869 & 0.886 & 0.920\\
Reghin & 0.651 & 0.880 & 0.861 & 1 & 0.886 & 0.983 & 0.845\\
Ludus & 0.729 & 0.993 & 0.867 & 0.996 & 1 & 0.784 & 0.845\\
Blaj & 0.785 & 0.867 & 0.886 & 0.983 & 0.784 & 1 & 0.896\\
Dumbraveni & 0.741 & 0.858 & 0.920 & 0.845 & 0.845 & 0.896 & 1\\ \bottomrule
\end{tabular}
\end{table*}
 
\subsection{Methods}\label{sec:methods}
\subsubsection{Environment and~Techniques} \label{sec:environment}
 the~chosen tool for~designing and~modeling the~data mining processes is~Rapid Miner~\cite{land2012rapid}. As~Hofmann and~Klinkenberg emphasized~\cite{hofmann2016rapidminer}, beside offering an almost comprehensive set of operators, it~also provides structures that express the~control flow for~a~process, in~a~presentation that is easy to~understand and~apply.
 
 Time series forecasting is~the~process of using a~model to~generate predictions for~future events based on known past events~\cite{kumar2014time}.
In~\cite{li2019short} a~wind speed forecasting is~based on an improved ant colony algorithm, as~ant-models are~used to~solve complex problem~\cite{complexproblem}; ant-models solve data mining tasks as clustering, classification and~prediction~\cite{nayak2020ant, azzag2006data}.

{To predict the~soil humidity for~a~location, the~time windowing technique was applied on the~source data.}
Koskela~et~al.~\cite{koskela1998time} specify that windowing is~used to~split the~time series into input~vectors. By~this approach, the~problem is~converted into selecting the~length and~type of window that will be~used. In~predicting the~soil humidity on a~specific date and~for a~specific location, the~machine learning algorithms use a~”window” of previous days~values. 

\begin{figure}[htbp]  
 	\centering
 	\includegraphics[scale=0.55]{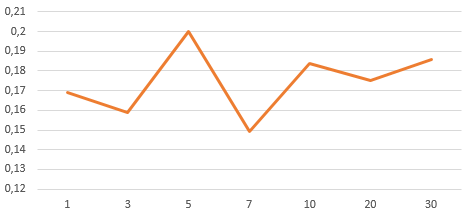}
 	\caption{Relative Error (RE) representation for~the~initial testing phase during various time windows (from 1 to~30 days): time windows (\emph{X}-axis) and~the~obtained RE values (\emph{Y}-axis).}
 	\label{fig:win_re}
 \end{figure}

In the~beginning of the~experiments, we~tried different values for~the~window considered, starting from one day, to~one week and~until one month worth of data (1, 3, 5, 7, 10, 20, 30). These first relative errors results for~various time windows are~depicted in~Figure~\ref{fig:win_re}. The~best results on our data were obtained using 7 days upfront~information.

The tests were performed using 80\% data for~creating and~training the~model and~20\% data for~validation.

\subsubsection{Machine Learning~Algorithms} \label{sec:mlAlg}
For investigating the~behaviour of the~results and~the~efficiency of the~proposed hybrid technique, more algorithms were~chosen:

\begin{itemize}
    \item \textbf{k-Nearest Neighbour (k-NN)}---as Cunningham and~Delany~\cite{cunningham2007k} mentioned, it~is~one of the~most straightforward machine learning techniques; 

   \item \textbf{Deep Learning (DL)}---not yet used in~the~industry as a~valuable option, even though deep learning had very successful applications in~the~last years~\cite{fawaz2019deep};
   \item \textbf{Gradient Boosted Trees (GBT)}---Yu~et~al.~\cite{yu2018data} used GBT to~predict the~short-term wind speed;
   \item \textbf{Decision Trees (DT)}---according to~Geurts~\cite{geurts2002contributions}, this algorithm is~``fast, immune to~outliers, resistant to~irrelevant variables, insensitive to~variable rescaling''.
\end{itemize}

These algorithms cover more or less all types of machine learning approaches, considering that:

\begin{itemize}[leftmargin=*,labelsep=5.8pt]
\item[-] k-NN is~a~straight forward and~most used mathematical model;

\item[-] Deep Learning means complex neural networks with advanced mathematics behind them;

\item[-] Gradient boosted trees represent a~mathematical approach to~decision trees;

\item[-] Decision trees are~algorithm-based discrete~models.
\end{itemize}

\medskip

The values for~the~algorithm's parameters were decided after running the~Optimize Parameter operator on various combinations, in~Rapid Miner. The~setup was then decided from the~values that produced the~best results in~terms of relative~error.

\begin{figure}[htbp]
 	\centering
 	\includegraphics[scale=0.6]{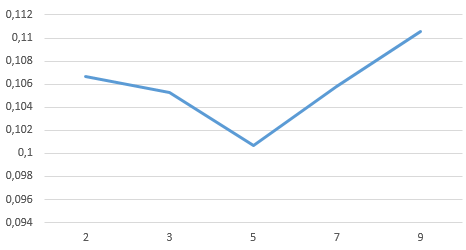}
 	\caption{Relative Error (RE) representation for~k-Nearest Neighbors (k-NN) during the~optimization parameter process: values tested for~$k$ (\emph{X}-axis) and~the~obtained RE values (\emph{Y}-axis).}
 	\label{fig:k_re}
 \end{figure}

Figure~\ref{fig:k_re} presents an overview of the~tests performed for~k-NN, for~ different values for~k. The~smallest RE were obtained when k was~5.
 
 \medskip
 
\begin{figure}[htbp] 
 	\centering
 	\includegraphics[scale=0.5]{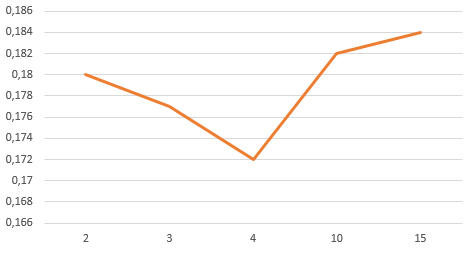}
 	\caption{Relative Error (RE) representation for~Decision Trees (DT) during the~optimization parameters process: the~values tested for~DT maximal depth (\emph{X}-axis) and~the~RE obtained values (\emph{Y}-axis).}
 	\label{fig:dt_re}
 \end{figure}

The optimization process with respect to~the~depth of the~decision trees has led us to~a~maximal depth of 4. Figure~\ref{fig:dt_re} shows the~relative error for~various depths. Table~\ref{table:dlOptimizeValues} includes the~parameter value combinations tested for~DL. Highlighted is~the~combination that provided the~lowest~error.

 For GBT we~tested the~results for~the~following combinations of values: number of trees ---from 10 to~100 with a~step of 10; maximal depth--values 3, 5, 7, 15; learning rate--- values 0.01, 0.02, 0.03, 0.1; number of bins---values 10, 20, 30. The~combination that performed best for~GBT, providing a~relative error of 0.143873273 is~depicted in~Table~\ref{table:parametersML}. 

Table~\ref{table:parametersML} presents the~settings used for~the~machine learning algorithms. This setup was the~same for~all scenarios that were studied, in~order to~have a~common point of reference when performing the~comparison for~the~results in~each described~scenario.
 
\begin{table*}[htbp]
\caption{Relative Error (RE) results for~Deep Learning (DL) in~the~parameter optimization~process.}
\centering
\label{table:dlOptimizeValues}
\begin{tabular}{cccccc}
\toprule
\multicolumn{2}{c}{\bf Activation: Tanh} &
\multicolumn{2}{c}{\bf Activation: Rectifier} &
\multicolumn{2}{c}{\bf Activation: ExpRectifier}
\\\midrule
{\bf Epochs} & \textbf{RE} & \textbf{Epochs} & \textbf{RE} & \textbf{Epochs} & \textbf{RE} \\
\midrule
2  & 0.163572675 &      2   & 0.135942805       & 2 & 0.145482752 \\
4  & 0.158022918 &      4   & 0.146780326       & 4 & 0.175774121 \\
6  & 0.157398315 &      6   & 0.182660829       & 6 & 0.172397822 \\
8  & 0.174560711 &      8   & 0.192005928       & 8 & 0.184494165 \\
10 & 0.159373990 & {\bf {10}} & {\bf {0.121232879}} & 10& 0.136397852 \\
15 & 0.175305445 &      15  & 0.186658629       & 15& 0.173097985 \\ 
\bottomrule
\end{tabular}
\end{table*}
\unskip

\begin{table*}[htbp]
\caption{Machine learning algorithms parameters~setting.} 
\centering
\label{table:parametersML}
\begin{tabular}{ll ll ll ll}
\toprule
\multicolumn{2}{c}{\bf k-NN} & \multicolumn{2}{c}{\bf GBT} &
\multicolumn{2}{c}{\bf DT} & \multicolumn{2}{c}{\bf DL} \\ 
\midrule
k: & 5 & Number of Trees: & 50 & Maximal depth: & 4 & Activation: & Rectifier \\ 
Measure: & Euclidean  & Maximal depth: & 7 & Minimal gain: & 0.01 & Epochs: & 5 \\ 
& distance& Learning rate: & 0.01 & Minimal leaf size: & 2 &  &  \\ 
& & Number of bins: & 20 &  &  &  &  \\ \bottomrule
\end{tabular}
\end{table*}

\subsubsection{Measurements~Performed} \label{sec:measurements}

Rapid Miner offers a~large set of possible performance criteria and~statistics that can be~monitored.  From~this set, the~following ones were chosen in~our~experiments:

\begin{itemize}
\item Absolute Error (AE)---the~average absolute deviation of the~prediction from the~actual value. This~value is~used for~Mean Absolute Error which is~very common measure of forecast error in~time series analysis~\cite{hyndman2014forecasting}.
    \item Relative Error~(RE)--- the~average of the~absolute deviation of the~prediction from the~actual value divided by actual value~\cite{abramowitz1965handbook}.
    \item Root Mean Squared Error (RMSE)---the~standard deviation of the~residuals (prediction errors). It~is~calculated by finding the~square root of the~mean/average of the~square of all errors~\cite{hyndman2006another}:
    \[
    RMSE = \frac{1}{n}\sum{i=1}{n}{(p_i-d_i)^2},
    \]
    where $n$ is~the~number of outputs, $p_i$ is~the~$i$-th actual output and~$d_i$ is~the~$i$-th desired output.
    \item Spearman $\rho$---computes the~rank correlation between the~actual and~predicted values~\cite{ref1}.
\end{itemize}

\subsection{Implementation}\label{sec:implementation}

For easier access, data used to~validate the~proposed technique, were saved in~a~local Rapid Miner~repository. 
For each location from the~six chosen, we~had available the~following information: date, average air temperature per day (centigrades) and~soil~humidity.

To validate the~proposed technique and~have as many variations as possible, more scenarios have been considered, starting from the~available data. The~value that was chosen to~be predicted was the~soil humidity for~a~specific location. 

The~air temperature was considered to~be the~contextual data for~the~scenario involving context-awareness. The~reason this qualified better as context is~because it~is~an information that can be~obtained from different sources, like sensors or other weather channels; it~can be~mined and~provide information on its own. As~correlated sources were chosen the~locations in~the~closest proximity with the~information on the~soil moisture~data.

In a~real life scenario there could be~more information available for~context/correlated sources, as~it was described in~Section~\ref{sec:analysis_technique}. For~the purpose of validating the~proposed technique, the~number of attributes used was minimized to~be able to~focus on the~implementation and~obtained~results. 

\medskip

The following scenarios served as basis for~our~research:

\medskip

\begin{itemize}
    \item{\bf Standalone}---predict the~soil humidity for~a~location, knowing previous evolution of the~soil humidity for~that location (main data).
    \item{\bf CADM}---predict the~soil humidity for~a~location, knowing:
                    previous evolution of the~soil humidity for~that location (main data); air temperature evolution for~the~location (context data).
    \item \textbf{CADM + CDM 1 source}---predict the~soil humidity for~a~location, knowing:
                    previous evolution of the~soil humidity for~that location (main data); 
                    air temperature evolution for~the~location (context data); 
                    soil humidity information for~one of the~closest locations (correlated source 1 data).
    \item \textbf{CADM + CDM 2 sources}---predict the~soil humidity for~a~location, knowing: previous evolution of the~soil humidity for~that location (main data); 
                air temperature evolution for~the~location (context data); 
                    soil humidity information for~two of the~closest locations (correlated source 1 data and~correlated source 2 data).
    \item \textbf{CADM + CDM 3 sources}---predict the~soil humidity for~a~location, knowing:
                    previous evolution of the~soil humidity for~that location (main data); 
                    air temperature evolution for~the~location (context data); 
                    soil humidity information for~three of the~closest locations (correlated source 1 data, correlated source 2 data and~correlated source 3 data).
    \item \textbf{CDM 3 sources}---predict the~soil humidity for~a~location, knowing:
                    previous evolution of the~soil humidity for~that location (main data); 
                    soil humidity information for~three of the~closest locations (correlated source 1 data, correlated source 2 data and~correlated source 3 data).
\end{itemize}

The described scenarios were used for~all locations and~all chosen machine learning~algorithms.

Table~\ref{tab:testScenariosExperiment} presents examples of the~combinations that served as study in~the~experiment for~predicting the~soil moisture for~two locations. Similar scenarios were run for~the~other four locations investigated in~Transylvania and~for the~ones in~Canada. The~question marks represent missing~values.

\begin{table*}[htbp]
\caption{Example of combined test scenarios used in~the~experiments. Notations: H\_Location and~T\_Location denotes the~humidity (H) and~respectively the~temperature (T)  of the~specified~location.}
\label{tab:testScenariosExperiment}
\centering
\begin{tabular}{cccccc}
\toprule
\textbf{Predicted } & \textbf{Context} & \textbf{ \begin{tabular}[c]{@{}l@{}}Correlated\\  Source 1\end{tabular}} & \textbf{\begin{tabular}[c]{@{}l@{}}Correlated\\  Source 2\end{tabular}} & \textbf{\begin{tabular}[c]{@{}l@{}}Correlated\\Source 3\end{tabular}} & \textbf{Scenario}\\ 
\midrule
H\_Sarmasu & T\_Sarmasu & H\_Reghin & H\_TMures & H\_Ludus & CADM+CDM 3~sources\\ 
H\_Sarmasu & T\_Sarmasu & H\_Reghin & H\_TMures & ? & CAD+CDM 2~sources \\ 
H\_Sarmasu & T\_Sarmasu & H\_Reghin & ? & ? & CADM+CDM 1~source \\ 
H\_Sarmasu & T\_Sarmasu & ? & ? & ? & CADM\\ 
H\_Sarmasu & ? & H\_Reghin & H\_TMures & H\_Ludus & CDM 3~sources\\ 
H\_Sarmasu & ? & ? & ? & ? & Standalone\\ 
H\_TMures & T\_TMures & H\_Reghin & H\_Sarmasu & H\_Ludus & CADM+CDM 3~sources \\ 
H\_TMures & T\_TMures & H\_Reghin & H\_Sarmasu & ? & CADM+CDM 2~sources\\
H\_TMures & T\_TMures & H\_Reghin & ? & ?  & CADM+CDM 1~source \\ 
H\_TMures & T\_TMures & ? & ? & ?  & CADM\\ 
H\_TMures & ? & H\_Reghin & H\_Sarmasu & H\_Ludus   & CDM 3~sources \\ 
H\_TMures & ? & ? & ? & ?  & Standalone \\ \bottomrule
\end{tabular}
\end{table*}

 For~each machine learning algorithm an adaptable Rapid Miner process was designed, as~described in~Figure~\ref{fig:flowChart}, that loaded the~test scenarios as designed, and~ran the~analysis based on the~setup of each scenario, registering the~results in~a~final repository. 
Section~\ref{sec:results} presents an overview of the~obtained results and~analysis.

\section{Results}\label{sec:results}

The Rapid Miner processes stored the~results for~the~measurements performed on the~accuracy of the~prediction (RE, RMSE, AE) in~the~format---value, standard deviation and~variance for~each~measure. 

\subsection{Overall Statistical~Results}

{An important issue of the~research was the~resilience of the~outcome relative to~the~various data sources and~inputs.} Therefore Spearman $\rho$~\cite{schmid2007multivariate} analysis was performed. 

Spearman $\rho$ is~a~non-parametric test used to~measure the~strength of association between two variables, where the~value $r = 1$ means a~perfect positive correlation and~the~value $r = -1$ means a~perfect negative correlation. Further~on, we~present the~conclusions based on the~study developed on the~analysis performed on RE and~Spearman $\rho$.

\begin{figure*}[htbp]
 	\centering
 	\includegraphics[scale=0.55]{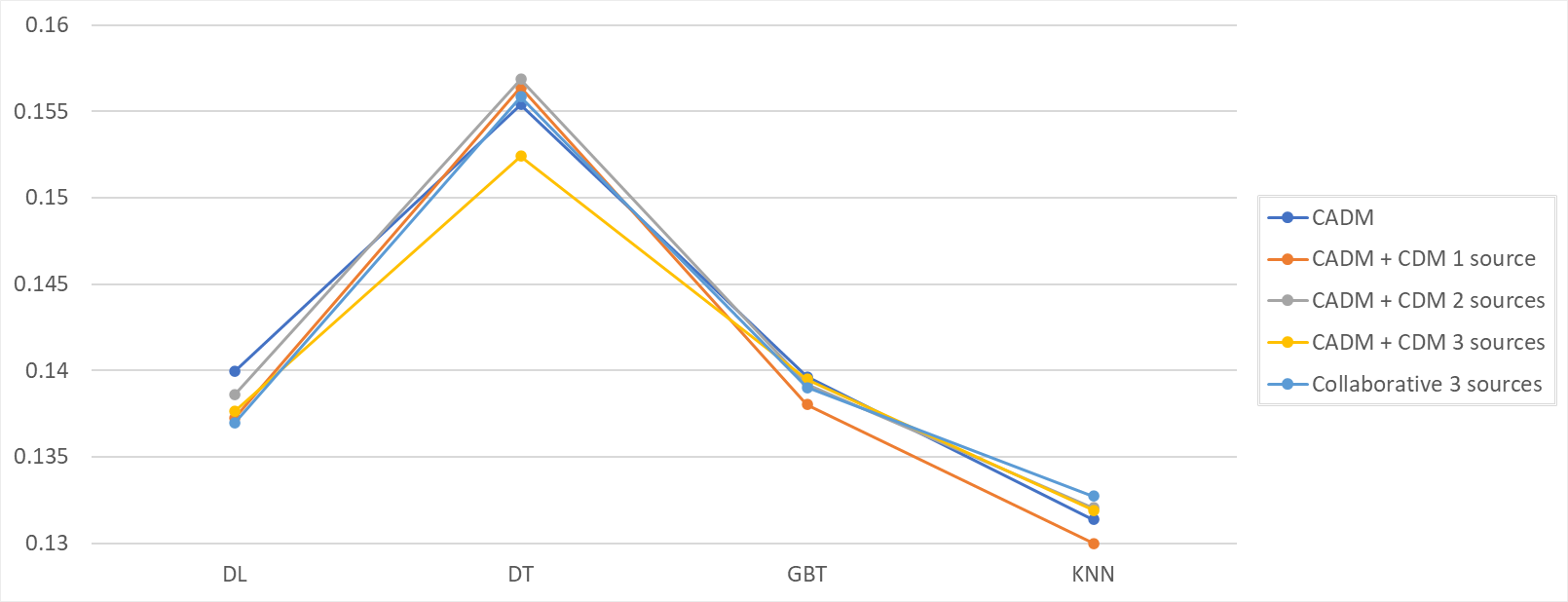}
 	\caption{Relative Error (RE) representation of the~overall results grouped by the tested algorithm: the~algorithms (\emph{X}-axis) and~the~RE obtained values (\emph{Y}-axis).}
 	\label{fig:rePerAlg}
 \end{figure*}

Figure~\ref{fig:rePerAlg} displays a~high level summarized overview of the~relative error for~all the~locations in~the~ Transylvanian data~source.

 Table~\ref{table:spearmanResults} presents an overview of the~obtained values for~the~Spearman $\rho$ coefficient, computed for~all the~algorithms and~scenarios, for~both data sources~\cite{russianPrognosis,albertaWeather} investigated, so that we~can check if~the~conclusions still stand in~a~different setup.

\begin{figure*}[htbp]
 	\centering
 	\includegraphics[scale=0.4]{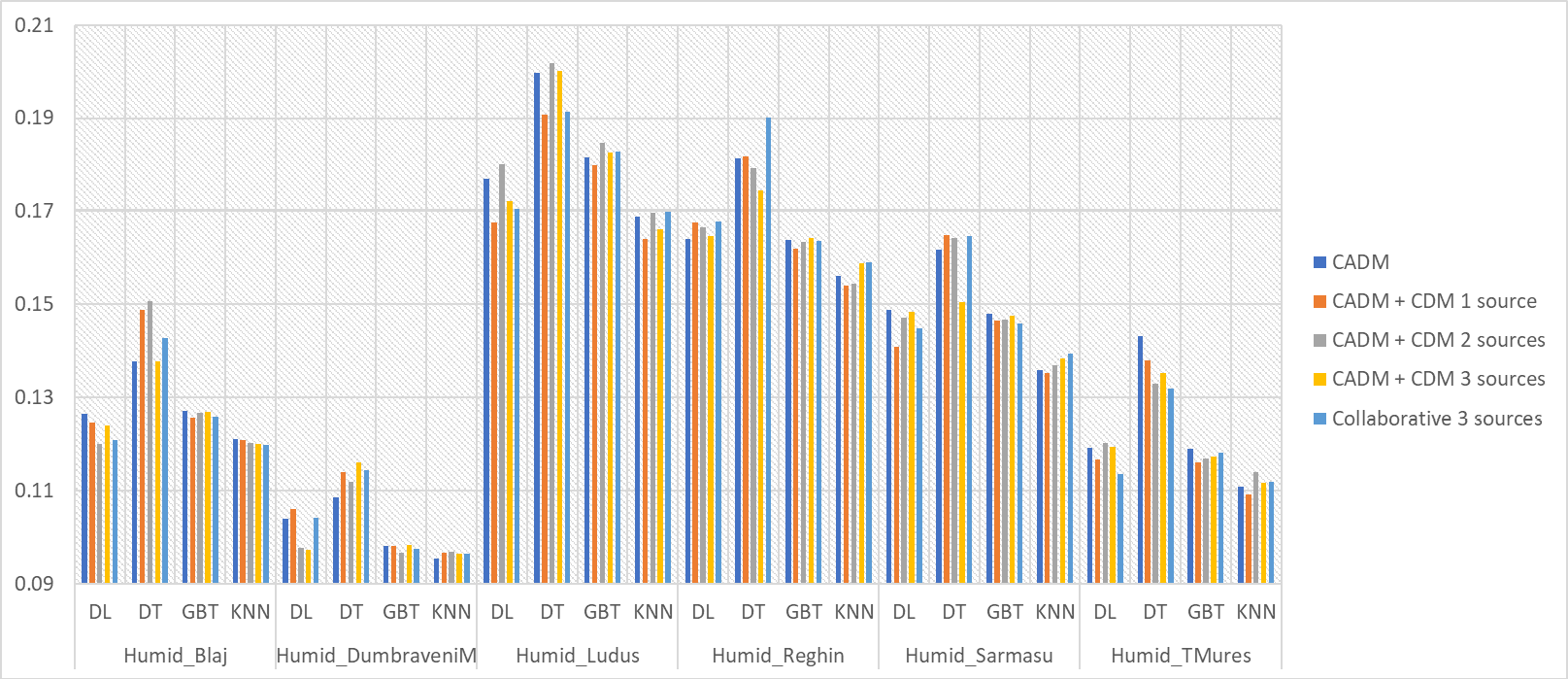}
 	\caption{{Relative Error (RE) overview representation per location and~algorithm:  location and~algorithm tested (\emph{X}-axis) and~the~RE obtained values (\emph{Y}-axis).}}
 	\label{fig:reOverview}
 \end{figure*}

Figure~\ref{fig:reOverview} displays a more specific overview of the relative error for each location and algorithm. Several discussion and conclusions follows.

\medskip

\begin{itemize}
     \item k-NN, overall, has the~smallest relative error, and~it is~a~solid candidate when choosing a~data mining technique, no matter the~chosen scenario. The~Spearman $\rho$ coefficient also provides the~best results for~both Canadian and~Transylvanian data source when using k-NN.
     \item GBT offers a~similar performance for~all scenarios in~terms of RE.
     \item Overall, for~DT, both the~RE report and~the~raking statistics show that the~best results are~obtained in~the~CADM + CDM 3 sources scenario and~in the~Collaborative with 3 sources scenario, emphasizing once again that the~combination of the~quality context data and~collaborative sources available, would improve the~results.
     \item for~DL, the~best result is~obtained also in~the~CDM + 3 sources scenario from the~RE perspective, but~from the~Spearman $\rho$ perspective, it~proves that the~data sources might influence the~results.
 \end{itemize}
 
\begin{table*}[htbp]
 \centering
 \caption{Spearman $\rho$ overall results for~Data Sources~\cite{russianPrognosis,albertaWeather} described in~Table~\ref{table:dataSources} and~tested~scenarios.}
 \label{table:spearmanResults}
\begin{tabular}{cccccc}
\toprule
\textbf{Data Source} & \textbf{Scenario} & \textbf{DL} & \textbf{DT} & \textbf{GBT} & \textbf{kNN} \\
\midrule
Transylvania & CADM & 0.80982 & 0.74204 & 0.87051 & 0.81566 \\
Transylvania & CADM + CDM 1 source & 0.84593 & 0.73765 & 0.88172 & 0.82123 \\
Transylvania & CADM + CDM 2 sources & 0.85932 & 0.75500 & \textbf{0.89184} & 0.81689 \\
Transylvania & CADM + CDM 3 sources & \textbf{0.86358} & \textbf{0.76217} & 0.87103 & \textbf{0.83141} \\
Transylvania & Standalone & 0.82657 & 0.72264 & 0.87448 & 0.81372 \\
Transylvania & CDM + 3 sources & 0.83730 & 0.76077 & 0.87631 & 0.81345 \\
\midrule
Canada & CADM & 0.61548 & 0.83449 & 0.87627 & 0.87236 \\
Canada & CADM + CDM 1 source & \textbf{0.73143} & 0.83447 & 0.87627 & 0.87513 \\
Canada & CADM + CDM 2 sources & 0.66200 & 0.83450 & 0.87412 & 0.87429 \\
Canada & CADM + CDM 3 sources & 0.70211 & \textbf{0.89523} & 0.86505 & \textbf{0.90480} \\
Canada & Standalone & 0.72011 & 0.80276 & 0.87412 & 0.86220 \\
Canada & CDM + 3 sources & 0.66265 & 0.83450 & \textbf{0.89818} & 0.88124 \\ \bottomrule
\end{tabular}
\end{table*}

\bigskip\bigskip

Nevertheless, the~study also shows that there might be~variations in~the~value of the~RE per each location, meaning that for~some locations, the~user might decide that the~best scenario is~the~CADM + CDM 1 source (e.g., for~DL and~Ludus because the~RE in~that specific case is~the~lowest); overall, the~CADM + CDM 3 sources or CDM 3 sources give the~best results. One could statistically decide, based on the~need at hand and what would be~the~best combination to~use in~a~specific~situation.

\subsection{Specific Scenario~Results}

\begin{figure*}[htbp]
 	\centering
 	\includegraphics[scale=0.6]{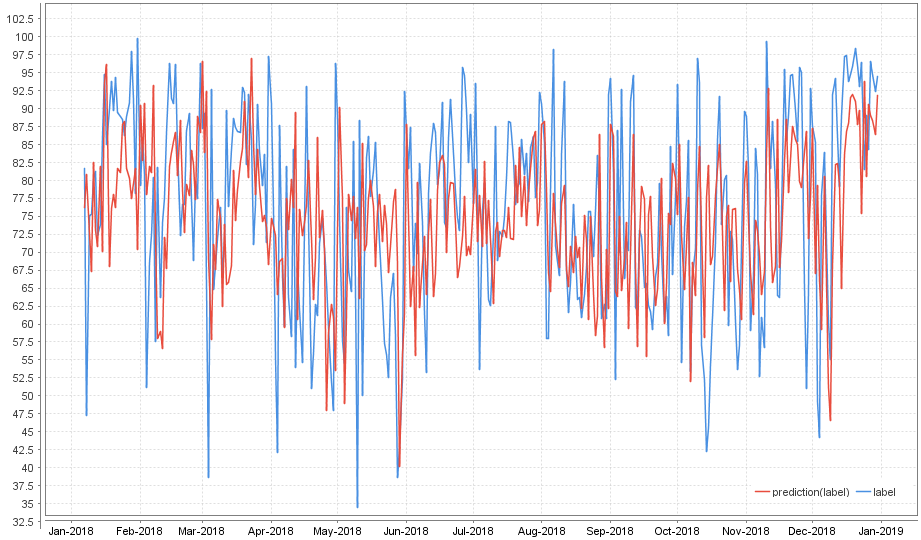}
 	\caption{Deep Learning (DL) prediction overview for~Sarmasu: CADM + CDM 3 sources scenario: the~values for~the~actual value, in~blue, and~predicted value, in~red (\emph{X}-axis) and~the~time series for~which the~results were registered (\emph{Y}-axis).}
 	\label{fig:dlOverviewLocation}
 \end{figure*}

\begin{figure*}[htbp]
 	\centering
 	\includegraphics[scale=0.55]{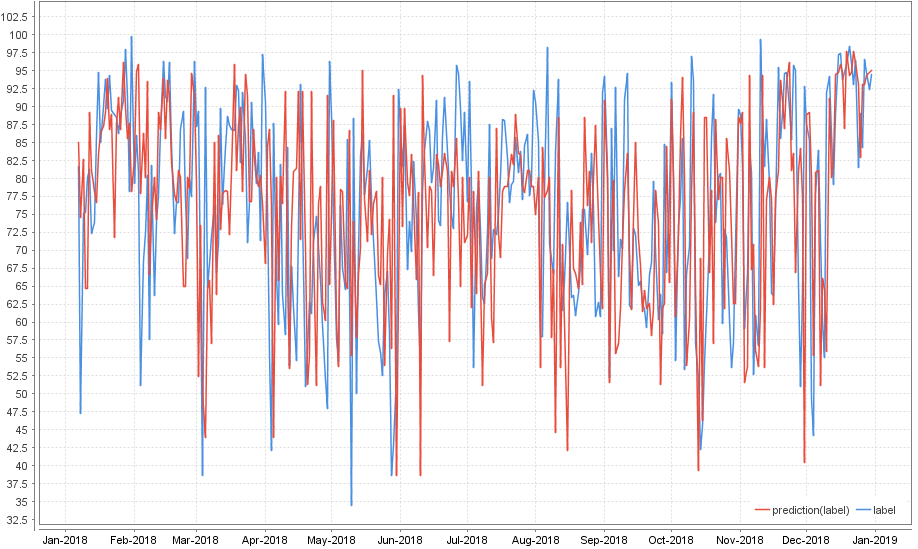}
 	\caption{Decision Tree (DT) prediction overview for~Sarmasu: CADM + CDM 3 sources scenario: the~values for~the~actual value, in~blue, and~predicted value, in~red (\emph{X}-axis) and~the~time series for~which the~results were registered (\emph{Y}-axis). }
 	\label{fig:dtOverviewLocation}
 \end{figure*}

\begin{figure*}[htbp]
 	\centering
 	\includegraphics[scale=0.55]{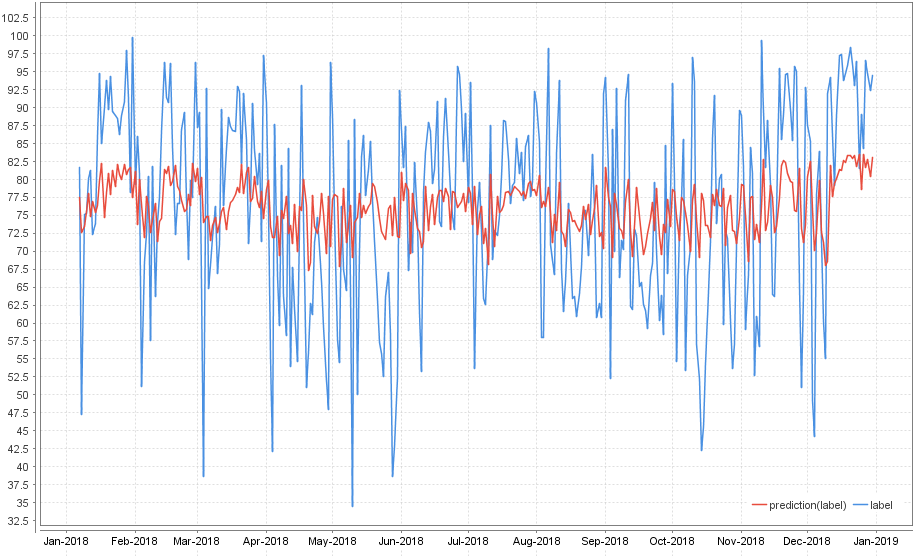}
 	\caption{Gradient Boosted Tree (GBT) prediction overview for~Sarmasu: CADM + CDM 3 sources scenario: the~values for~the~actual value, in~blue, and~predicted value, in~red (\emph{X}-axis) and~the~time series for~which the~results were registered (\emph{Y}-axis).}
 	\label{fig:gbtOverviewLocation}
 \end{figure*}

\begin{figure*}[htbp]
 	\centering
 	\includegraphics[scale=0.55]{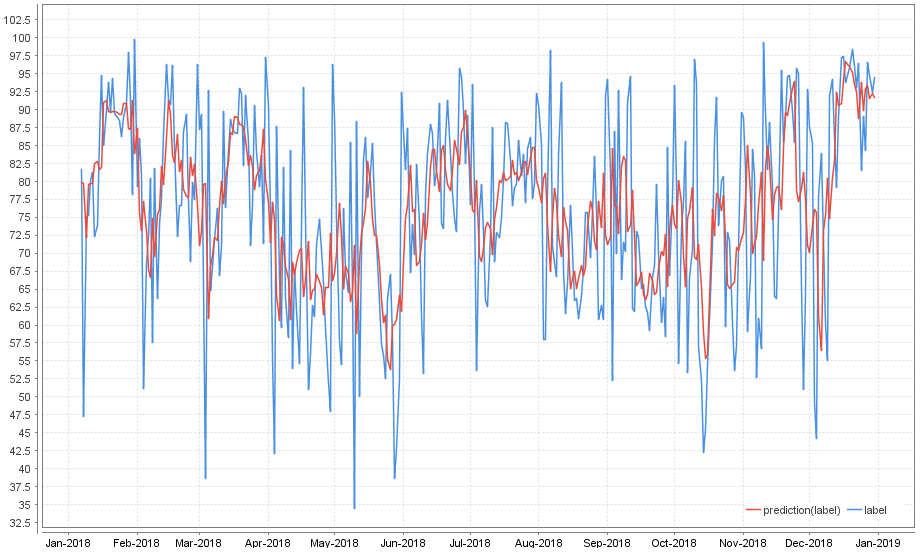}
 	\caption{k-Nearest Neighbor (k-NN) prediction overview for~Sarmasu: CADM + CDM 3 sources scenario: the~values for~the~actual value, in~blue, and~predicted value, in~red (\emph{X}-axis) and~the~time series for~which the~results were registered (\emph{Y}-axis). }
 	\label{fig:knnOverviewLocation}
 \end{figure*}

A deeper analysis can be~performed for~a~specific location, for~each candidate scenario and~algorithm, to~understand the~way the~prediction fluctuates versus the~actual value. For~example, for~the test scenario CADM + CDM 3 sources, for~a~specific location (Sarmasu) we~could check the~graphical overview of the~variations of the~predictions for~each algorithm studied. Figure~\ref{fig:dlOverviewLocation} offers the~overview for~the~DL algorithm, Figure~\ref{fig:dtOverviewLocation} for~the~DT, while Figures~\ref{fig:gbtOverviewLocation} and~\ref{fig:knnOverviewLocation} present the~overview for~GBT, respectively k-NN. In~blue is~the~graphical representation of the~soil humidity value, while in~red are~represented the~predicted~values. 

\unskip
\unskip
 
 \unskip

\begin{figure*}[htbp]
 	\centering
 	\includegraphics[scale=0.25]{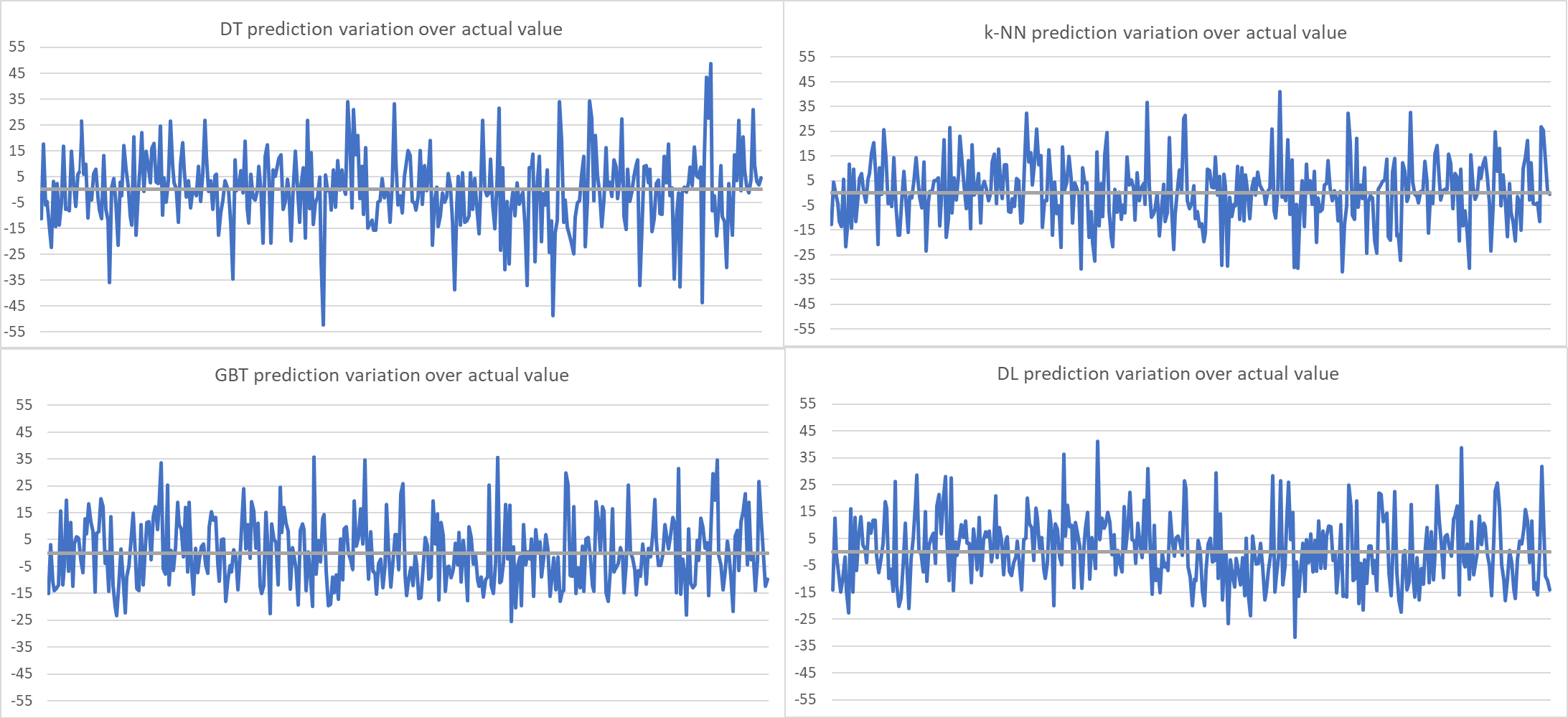}
 	\caption{Prediction variation overview for~Sarmasu: CADM + CDM 3 sources scenario for~DT \& k-NN (upper side) and~GBT \& DL (lower side) the~values for~the~deviation of predicted value versus actual value (\emph{X}-axis) and~the~time series for~which the~results were registered (\emph{Y}-axis). }
 	\label{fig:predOverviewLocation}
 \end{figure*}

Figure~\ref{fig:predOverviewLocation} depicts the~differences between actual and~predicted values for~all the~algorithms, while Table~\ref{table:SarmasuSD} presents the~standard deviation overview for~the~values~represented. 

It can be~observed that the~lowest deviation is~produced by the~GBT algorithm, but~if we~look at the~representation, it~can be~concluded that the~reason this happens is~because the~predicted value varies around the~average of the~actual value with the~chosen setup for~the~algorithm, making it~not a~valid option in~the~soil moisture prediction scenario, when one would expect predictions closer to~the~real value. Hence, the~best candidates for~the~problem are~k-NN and~DL~algorithms.

\begin{table}[htbp]
\centering
\caption{Standard deviation overview per algorithm for~Sarmasu in~the~CADM + CDM 3 sources~scenario.}
\label{table:SarmasuSD}
\begin{tabular}{cccc}
\toprule
\textbf{Alg} & \textbf{Value} & \textbf{Std.Dev.} & \textbf{Std.Dev.(\%)} \\ \midrule
\textbf{k-NN} & 0.138381577 & 0.026396991 & 19.08 \\ 
\textbf{DL} & 0.148357962 & 0.02705544 & 18.24 \\ 
\textbf{GBT} & 0.147488567 & 0.019781613 & 13.41 \\ 
\textbf{DT} & 0.150515614 & 0.033728874 & 22.41 \\ \bottomrule
\end{tabular}
\end{table}

\begin{figure*}[htbp]
 	\centering
 	\includegraphics[scale=0.7]{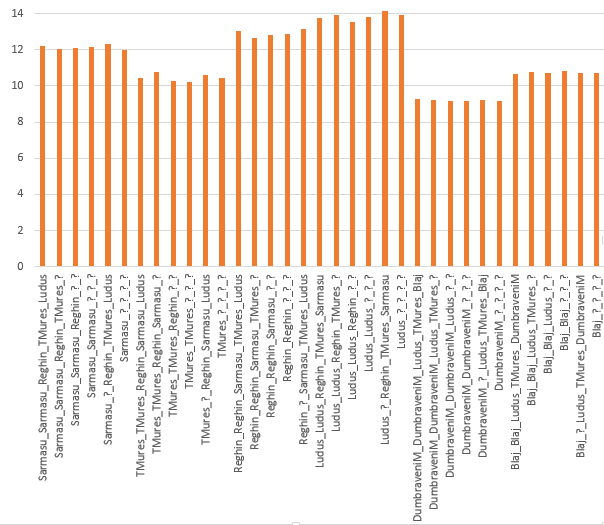}
 	\caption{The RMSEs for~k-NN algorithm with various setups:  \texttt{loc\_context\_colsrc1\_colsrc2\_colsrc3} (\emph{X}-axis) where \texttt{loc} is~the~location for~which the~prediction is~run, \texttt{context} is~the~contextual data for~that location, \texttt{colsrc1}, \texttt{colsrc2} and~\texttt{colsrc3} are~the~collaborative data sources, and~the value of RMSE (\emph{Y}-axis); $?$ denotes missing data~source.}
 	\label{fig:knnmse}
 \end{figure*}

\begin{figure*}[htbp]
 	\centering
 	\includegraphics[scale=0.7]{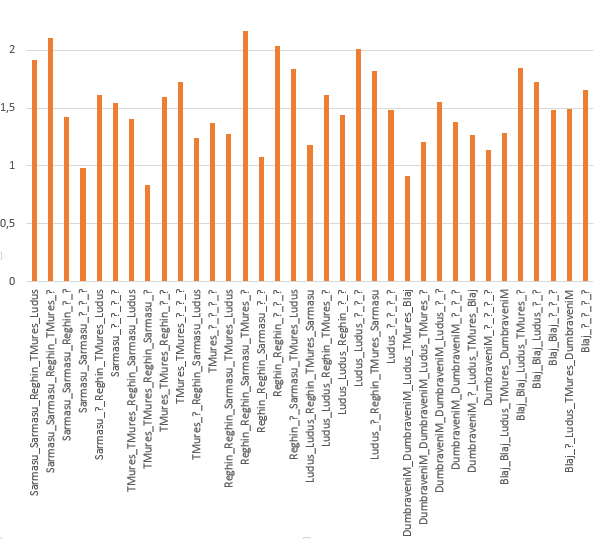}
 	\caption{The standard deviations for~k-NN algorithm with various setups: \texttt{loc\_context\_colsrc1\_colsrc2\_colsrc3} (\emph{X}-axis) where \texttt{loc} is~the~location for~which the~prediction is~run, \texttt{context} is~the~contextual data for~that location, \texttt{colsrc1}, \texttt{colsrc2} and~ \texttt{colsrc3} are~the~collaborative data sources, and~the value of standard deviation (\emph{Y}-axis); $?$ denotes missing data~source.}
 	\label{fig:knnstdev}
 \end{figure*}

 As k-NN has the~best performance, further we~present details of the~mean squared errors (MSE) (Figure~\ref{fig:knnmse}) and~the~standard deviations (in Figure~\ref{fig:knnstdev}) obtained using k-NN with various setups. 
In Figures~\ref{fig:knnmse} and \ref{fig:knnstdev}, the~\emph{X}-axis is~coded as \texttt{loc\_context\_colsrc1\_colsrc2\_colsrc3}, where \texttt{loc} is~the~location for~which the~prediction is~run, \texttt{context} is~the~contextual data for~that location, \texttt{colsrc1},  \texttt{colsrc2} and~ \texttt{colsrc3} are~the~collaborative data sources. When the~question mark appears, it~means that that data source is~missing.   

Figure~\ref{fig:knnmse} shows that the~highest errors occurs when there is~just one data source, respectively when there are~all of them---the~data from the~location at stake, the~context and~the~three collaborative data sources. In~the former case, the~high error is~due to~the~relatively low amount of data available, whereas in~the~latter one, the~error occurs from the~redundant quantity of data and~the~possible conflicts among them (as they are~not fully correlated, as~expected). The~best results (lowest errors) are~obtained when there are~two or three data~sources. 
 
\begin{figure*}[htbp]
 	\centering
 	\includegraphics[scale=0.9]{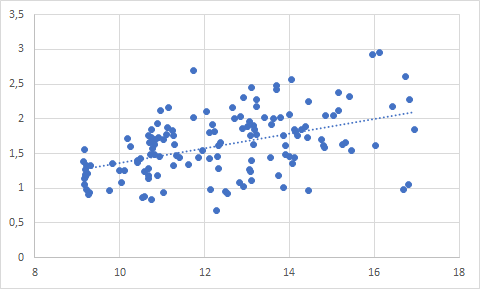}
 	\caption{The correlation between RMSE and~the~standard deviation: the~RMSE (\emph{X}-axis) and~the~standard deviation (\emph{Y}-axis).}
 	\label{fig:correl_mse_stdev}
 \end{figure*}

 An interesting point, shown in~Figure~\ref{fig:correl_mse_stdev}, is~that between the~RMSE and~the~standard deviations there is~a~very high correlation, of~0.953, which means that a~high error means more or less a~high standard deviation and~vice-versa.

\unskip

\section{Conclusions}\label{sec:conclusions}
Considering the~rapid increase of available data, no matter the~domain, finding improvements in~the~way data mining processes are~performed is~a~subject of continuous research. Previous work has shown the~advantages of using CADM and~CDM techniques over the~classic data mining process. The~current work presents the~basis of a~new technique for~combining the~two approaches in~a~flexible way that allows testing the~performance of different scenarios, easily configurable by the~user.

The technique was then applied on a~simple real life scenario for~predicting the~soil humidity for~more locations. Once again was proven that CADM and~CDM improve the~classical standalone results. The~algorithm with the~best overall results was k-NN, followed by DL. 

\bigskip

The~advantages of using the~proposed technique for~testing various CADM - CDM scenarios~are:

\medskip

\begin{itemize}
    \item the~possibility to~embed the~context of the~main data source;\smallskip
    \item the~possibility to~embed correlated data and~apply machine learning techniques on all of them;\smallskip
    \item allowing to~test multiple variations of scenarios in~a~single run, without~human intervention;\smallskip
    \item rapid introduction of a~new testing scenario, if~needed;\smallskip
    \item flexibility in~easily adding a~new machine learning algorithm to~be tested;\smallskip
    \item adding a~new attribute to~the~context or to~the~correlated source is~only a~configuration task, not influencing the~overall process.
\end{itemize}

\bigskip\bigskip

The described technique was thought and~tested in~the~CADM + CDM scenarios, because~testing various combinations was costly and~usually meant creating new processes for~each scenario. By~using the~new approach, it~changed in~a~configuration process. If~context and~collaborative sources are~not present, the~tested situation is~the~traditional data mining~process.

For now, the~current research focused on defining and~implementing a~flexible technique that would allow combining the~CADM and~CDM approaches in~various test scenarios, to~provide useful insights and~support for~deciding which is~the~best suitable approach for~a~specific real situation. As~this part was successfully covered, the~analysis of the~results is~yet a~step that had to~be performed and~was based mainly on the~experience of the~user. Considering this, further research might improve that part by defining important criteria that would make a~scenario the~best one for~forecasting. The~process could then be~improved by introducing this criteria and~make a~preliminary analysis of the~results by performing a~scoring on the~performance of each test scenario. Also statistical analysis of the~results could be~performed.

A step further on the~research would be~validating the~technique on larger, more complex data sets also from other domains of~interest.

\bigskip

\noindent{\bf Acknowledgment}
{This work has received funding from the~CHIST-ERA BDSI BIG-SMART-LOG and~UEFISCDI COFUND-CHIST-ERA-BIG-SMART-LOG Agreement no. 100/01.06.2019.}

\end{document}